\documentclass[page-classic]{epl2}
\usepackage{graphicx}

\title{Orbital magnetism in transition-metal systems: The role of local
correlation effects}
\author{S.~Chadov\inst{1} \and J.~Min\'{a}r\inst{1} \and
  M.~I.~Katsnelson\inst{2} \and H.~Ebert\inst{1} \and
  D.~K\"odderitzsch\inst{1} \and A.~I.~Lichtenstein\inst{3}
}
\shortauthor{S.~Chadov \etal}
 
\institute{                    
  \inst{1} Dept.~Chemie    und    Biochemie, 
Physikalische  Chemie,  Universit\"at  
M\"unchen,  Butenandtstr.   5-13, D-81377 M\"unchen,  Germany\\
  \inst{2} Institute for Molecules and Materials, Radboud University
  Nijmegen, NL-6525   ED  Nijmegen,   The  Netherlands\\
  \inst{3}Institute  of Theoretical  Physics, University  of Hamburg,
Germany
}
\pacs{71.15.Rf}{Relativistic effects}
\pacs{71.20.Be}{Transition metals and alloys}

\abstract{
The influence of  correlation effects on the orbital  moments for transition
metals  and  their alloys  is  studied  by first-principle  relativistic
Density Functional  Theory in combination with  the Dynamical Mean-Field
Theory.   In contrast  to  the  previous studies  based  on the  orbital
polarization  corrections  we  obtain  an improved  description  of  the
orbital moments for wide range of studied systems as bulk Fe, Co and Ni,
Fe-Co disordered alloys and 3$d$ impurities in Au. The proposed 
scheme can give simultaneously a correct dynamical description of the 
spectral function as well as static magnetic properties of correlated 
disordered metals.
}

\begin{document}

\maketitle

The  growing   interest  in  magnetic  materials,   their  surfaces  and
nanostructures requires improved theoretical first-principle methods for
their description,  in particular, when  a complex behavior  of magnetic
properties is observed  as in low dimensional systems  as, e.g. magnetic
clusters,    multilayers,   thin    films   and    magnetic   impurities
\cite{Sto99,GRV+03,BBB+05}.       Their      magnetic      anisotropies,
magneto-optical   spectra,  magnetic   dichroism  and   other  important
properties are  caused by spin-orbit coupling.  While  the spin magnetic
moments for 3$d$-transition metals (3$d$-TM), their alloys and impurities
in  non-magnetic  host  are   described  rather  accurately  by  density
functional theory  in the  local spin-density approximation  (LSDA), the
orbital moments are systematically  underestimated.  The reason for this
is  well-known: the  functional  variables of  the  LSDA potential  (the
charge  and  spin  density)   are  defined  as  averages  over  occupied
orbitals.  It  is  natural  that  such an  approximation  gives  a  good
description only for the quantities  which are slightly dependent on the
deviations  of  orbital  occupation   numbers  from  their  average,  as
e.g. spin magnetic moments.

An often used  approach to improve the description  of orbital magnetism
is the so-called orbital  polarization correction (OP) scheme introduced
by  Brooks {\it  et  al.}   \cite{Bro85,EJB89,EBJ90} in  a  form of  an
additional {\it  ad hoc}  term to  the Hamiltonian. As it was  shown by
Ebert and Battocletti \cite{EBG97} the OP enhancement of the orbital
moment partially could be realized by utilizing the more general current
density  functional theory.   Analyzing the  CDFT Eschrig  {\it  et al.}
\cite{ESKR05} have derived a systematic expression for the OP correction
(for    an    overview    and    results   of    this    approach    see
Ref.~\cite{Sar06}).    However,  despite   of  a   quite  accurate
description of  the orbital  moments in pure  3$d$-TMs and  their alloys
\cite{SEJ+92,EB96}, the LSDA+OP calculations noticeably overestimate the
unquenched  orbital moments  of the  3$d$-TM impurities  in  noble metal
hosts \cite{BSS+04,Fro04,Sar06}.   In the case of  clusters deposited on
metal surfaces Gambardella~{\em  et~al.}~\cite{GRV+03} have noticed that
the Racah parameter has to be reduced by about 50\% in order to describe
the experimental  orbital moments correctly.  However, in  this case the
calculated    magnetic    anisotropy   is    still    much   too    high
\cite{GRV+03,NCZ+01}.

An alternative approach  is based on the explicit  account for the local
(on-site)   many-body   correlations.    In  particular,   Solovyev~{\em
et~al.}~\cite{SLT98,Sol05}  have  shown  on  the basis  of  calculations
within the random-phase approximation that  the OP picture is one of the
limits of  the more general LSDA+U concept  \cite{AZA91} and the
later can  provide a better agreement  with experiment for  pure Fe, Co
and Ni.

On  the  other  hand,  the  LSDA+U  approach  fails  to  give  a  proper
description for the spectral  properties of the 3$d$-TMs having problems
with  the bandwidth,  spin splitting  and  satellite in  Ni, absence  of
quasiparticle damping, etc. \cite{LKK01,GMK+07}.

A state of the art way  to treat the local correlation (Hubbard) effects
is   based  on  the   Dynamical  Mean-Field   Theory  (for   review  see
\cite{KSH+06}) which  takes into account  dynamical correlation effects,
in  particular  spin-flip processes  induced by  fluctuations.  Combined
with the  LSDA this  scheme (LSDA+DMFT) provides  a very  reliable basis
explaining  a wide  range  of  both static  and  spectral properties  of
3$d$-TM  materials (magneto-optics,  photoemission, total  energy, etc.)
\cite{LKK01,PCE03,MEN+05,BME+06,GMK+07}.    In  contrast   to   all  the
previous approaches this method allows to study systematically the temperature
dependence of the electronic structure and gives an adequate description
of  magnetic properties of  Fe and  Ni in  a broad  temperature range
\cite{LKK01}.

The well-known complication  of combining the LSDA with  the DMFT is the
uncertainty in separating of the  Hubbard Hamiltonian from the LSDA one,
the so-called double-counting problem.   As it was indicated by numerous
DMFT studies, the static many-body  effects which can be overcounted in
the LSDA+DMFT  combination are relatively small  in 3$d$-TMs. Accordingly,
for the description of  spectral properties the established procedure is
to leave  only the dynamical  part of the  self-energy by setting  it to
zero at  the Fermi level \cite{LKK01,KL02,GMK+07}.  However,  as it was
recently shown  by Braun {\it et al.}  \cite{BME+06}, that this approximation
is  not  sufficient  for a  precise description  of  the  angular-resolved
photoemission   spectra  of   Ni  and   an  additional   static  orbital
polarization should be included.

In this letter  we demonstrate that accounting for the orbital polarization in
the static part of the LSDA+DMFT provides a proper description not only for
the  spectral properties  but also  for  the spin  and orbital  magnetic
moments for a wide range of 3$d$-TM systems (bcc Fe,  hcp Co and fcc
Ni, bcc  Fe$_x$Co$_{1-x}$ disordered alloys as well  as 3$d$-impurities in
the Au host).

The  calculations were done within  the  relativistic  full
potential Green's function (SPR-KKR)  method \cite{MCP+05}.   As a
DMFT-solver  the relativistic  version of  the  so-called Spin-Polarized
T-Matrix     Plus    Fluctuation    Exchange     (SPTF)    approximation
\cite{KL02,PKL05} was used.  According to this scheme the local
Green's function is obtained by
the corresponding site projection of the full KKR Green's function. 
The local Green's function is needed to obtain the bath 
Green's function for the Anderson impurity model 
via the  saddle-point equation. The bath Green's function is used 
as an input for the SPTF scheme to calculate the
local self-energy. The latter is added as an additional
energy-dependent potential in the radial Dirac equation which is solved
to calculate the new full KKR Green's function. This procedure
is repeated until the self-consistency in both the self-energy and the charge
density is achieved. The scheme has already been successfully applied 
for the  description of magneto-optics \cite{CME+06} and photoemission 
\cite{MEN+05,BME+06} in 3$d$-TMs including corresponding matrix 
element effects. 

 In the 
present work we concentrate on a more accurate account for the orbital 
polarization when calculating the self-energy. This is achieved by treating 
the static part of the self-energy on a Hartree-Fock  level
(first-order contribution  in terms  of the Coulomb
interaction)  described  in a  local approximation  to 
the self-energy  by  the  LSDA+U  method. The  static
double-counting  is subtracted  from  the self-energy  in the  so-called
``around mean-field limit''(AMF) of LSDA+U as given by Czyzyk and Sawatsky
\cite{CS94}:
\begin{eqnarray}
  V_{m\sigma}^{\rm LSDA+U}(\vec r) - V^{\rm LSDA}_{m\sigma}(\vec r)
  \approx\sum_{m'}U_{mm'}\left(n_{m'-\sigma}-n_{-\sigma}^0\right) +\!\sum_{m'\ne
  m}\left(U_{mm'}-J_{mm'}\right)\left(n_{m'\sigma}-n_{\sigma}^0\right),
\end{eqnarray}
where $n_{m\sigma}$ are the occupation numbers of the localized $d$-orbitals
and $n^0_{\sigma}$ stands for the sum $\frac{1}{2l+1}\sum_{m}n_{m\sigma}$.

The self-energy within the DMFT can be calculated in terms of two
parameters: averaged screened Coulomb interaction $U$ and exchange
interaction $J$.
For the
latter the screening is usually not crucial; the value of $J$
can  be calculated  directly from  LSDA  and is  approximately equal  to
0.9~eV  for all 3$d$ elements. This  value has  been adopted  for all
calculations  presented  here.  Different  methods  of  calculating  the
screened Coulomb  interaction $U$  for 3$d$-TMs lead  to estimates  in the
range between 2-3~eV which have  been used here (see discussion below). 

A  comparison  to  experiment  for  the spin  and  orbital  magnetic
moments calculated  within LSDA+DMFT  for bcc Fe, hcp Co and fcc Ni using
the experimental values for the lattice parameters 
is shown in Fig.~1. The self-energy was parameterized using the 
values $U=3$~eV for Co and Ni, and $U=2$~eV for Fe.
\begin{figure}
\includegraphics[width=0.45\textwidth,angle=0]{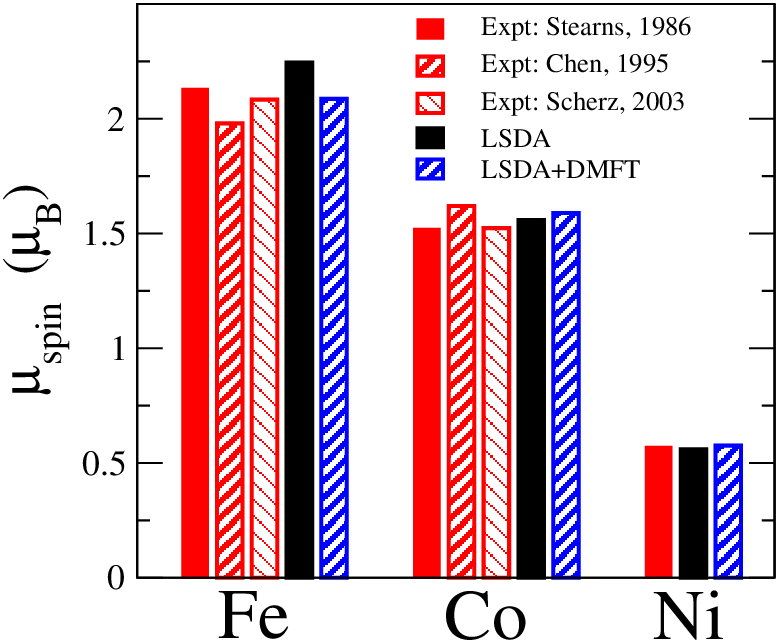}~
\includegraphics[width=0.45\textwidth,angle=0]{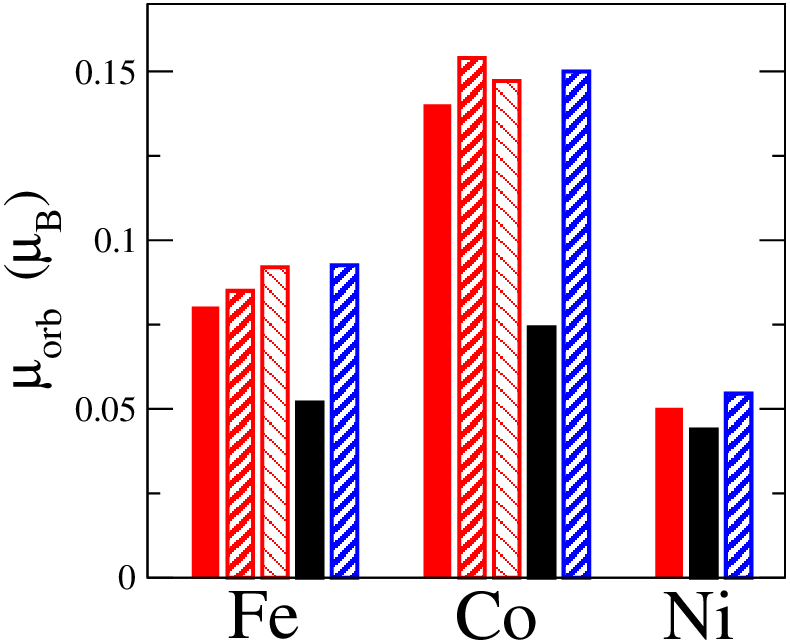}
\caption{(color online) Spin (left panel) and orbital (right panel)
  magnetic moments  in bcc Fe, hcp Co and fcc Ni  calculated using LSDA+DMFT 
(hatched blue bars) compared with plain LSDA calculations 
(black filled bars) and experimental data (red bars) 
taken from Refs.~\cite{Ste84,Sch03,CIL+95}. The corresponding DMFT parameters 
 are $U_{\rm Fe}=2$~eV, $U_{\rm Co}=U_{\rm Ni}=3$~eV 
and $J_{\rm Fe}=J_{\rm Co}=J_{\rm Ni}=0.9$~eV.}
\end{figure}
As expected, the  LSDA+DMFT approach  gives  results similar  to the  OP
correction: the small orbital splittings imposed by the LSDA+DMFT around
the  Fermi level  have almost  no effect  on the spin moment, but enhance the
orbital moment in an appreciable way.

By construction  the dynamical part of the self-energy $\Sigma$ 
in the vicinity of the Fermi level behaves like that of a Fermi liquid. 
Thus it cannot noticeably affect integral quantities
as spin and orbital magnetic moments. 
On the other hand, the applied AMF  static double 
counting which splits the orbitals only slightly 
at the Fermi level, has no impact on the renormalization of the density
of states. 
As it follows from Fig.~2, the total DOS curves calculated
within LSDA and LSDA+U as well as within LSDA+$\Sigma$ (e.g. only dynamical
part of self energy is used) and
LSDA+DMFT  are nearly indistinguishable. 

As the energy shifts of the $(-m,-m)$ and $(m,m)$ matrix elements
 of the Green's function occur in opposite directions, the total
DOS shift for a given spin character appears to be small.
As a result, the most affected quantity is the orbital
magnetic moment while the change of the spin moment is negligible. At the
same time the renormalization of the spectrum is controlled by
the dynamical part of the self-energy.
\begin{figure}
\includegraphics[width=0.9\textwidth,angle=0]{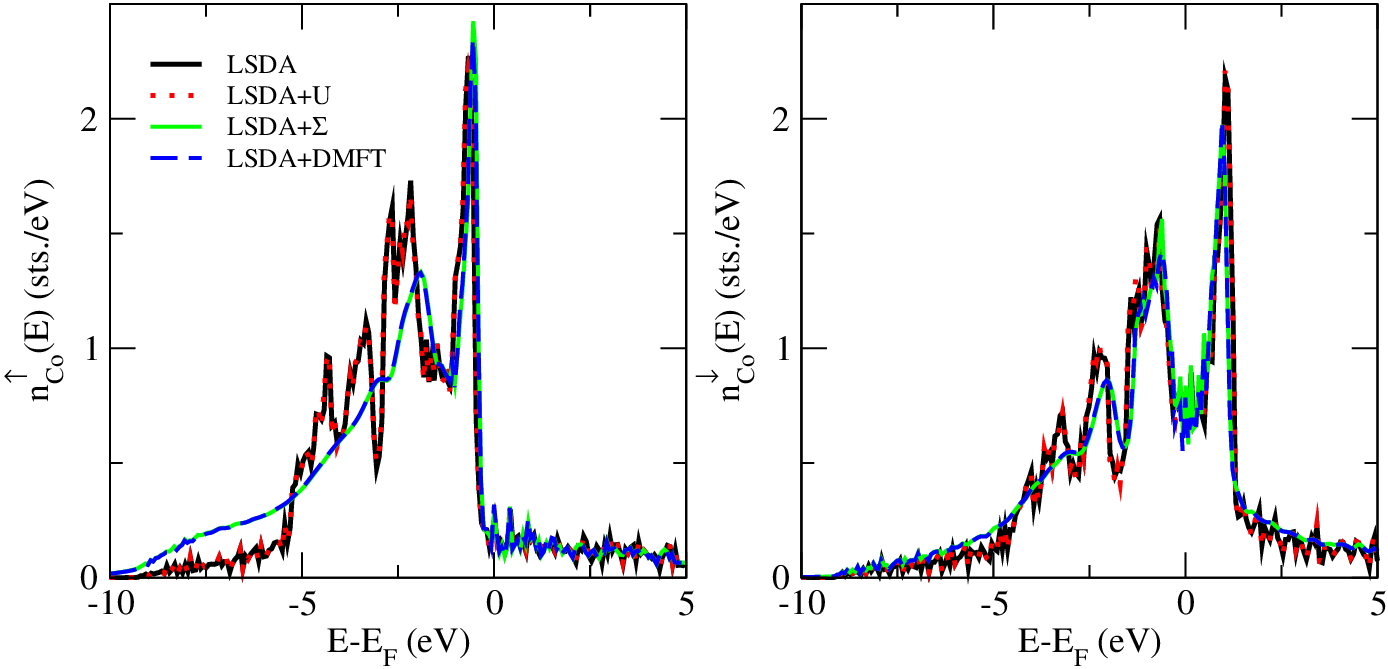}
\caption{(color online) Spin-resolved total DOS for hcp Co calculated
  via different approaches: black solid curve corresponds to plain LSDA,
  dotted (red) curve - accounts for local static correlations only
  (LSDA+U), light (green) solid line - only dynamical part of the
  self-energy (LSDA+$\Sigma$)   is used, dashed (blue) - both static and
  dynamic correlations are taken into account (LSDA+DMFT).
 The corresponding DMFT parameters are $U_{\rm Co}=3$~eV and 
 $J_{\rm Co}=0.9$~eV.}
\end{figure}



It follows from the various DMFT 
studies  as  well as  from  the  DMFT+GW-based calculation  \cite{BAG03}
that realistic values  of $U$ for 3$d$-TMs  are found between  2-3~eV.
As it is 
shown in  Fig.~1   this  range   of  $U$
parameters  brings  both  spin  and  orbital  moments  into  very  close
agreement with experiment.
 In the case of Fe the deviation of the orbital moment for $U$ 
above $2$~eV are found to be rather big (see Fig.~3), 
so that  the optimal values  of $U$ are confined  within 1.5-2~eV.
On the other hand, it was already proposed \cite{LKK01} 
that the local approximation (DMFT) works much better for Ni 
and Co than for Fe due to relative softness of magnons in the latter
case.  Recently, the  essential non-locality of correlation effects in
Fe was also demonstrated experimentally by angle-resolved 
photoemission \cite{SHR+05}.

\begin{figure}
\includegraphics[width=0.9\textwidth,angle=0]{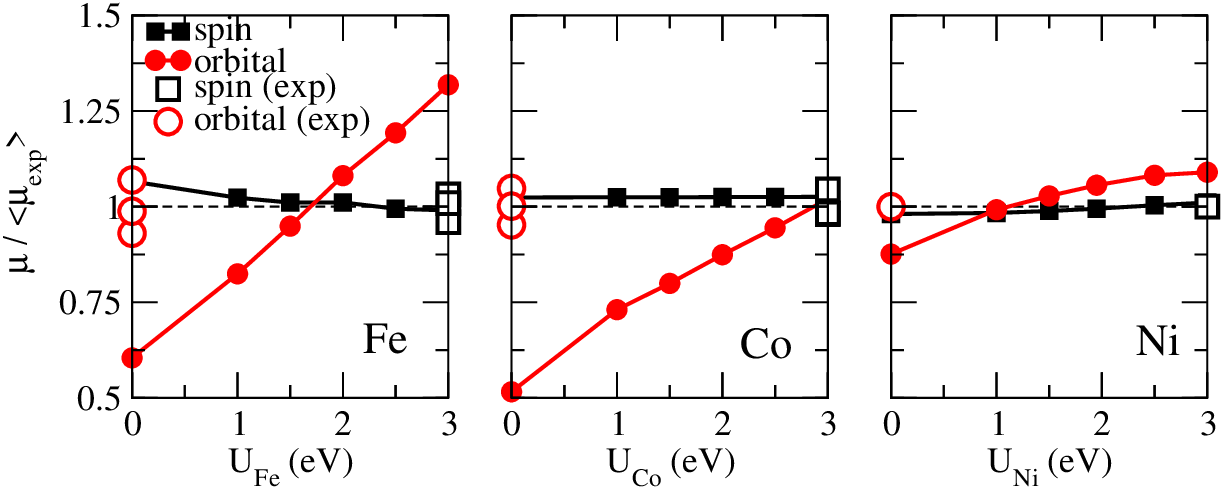}
\caption{(color online) Dependence of spin (black squares) and orbital
  (red circles) magnetic moments on $U$ parameter. Magnetic moments are
  given in units of ratios of $<\mu_{\rm exp}>$ which is 
  the average over the experimental moments (taken from
  Refs.~\cite{Ste84,Sch03,CIL+95}). Experimental values are marked
  with open squares (spin) and circles (orbital). The case $U=0$
  corresponds to the plain LSDA calculations. For all $U\ne 0$ the $J$
  parameter is fixed to 0.9~eV.
}
\end{figure}


It is worth to mention that some of the experimental results
(Refs.~\cite{Sch03,CIL+95}) are obtained from the measurements of the
x-ray magnetic circular dichroism (XMCD) at the $L_{2,3}$-edges. 
The sum rules used to derive the spin and orbital
magnetic moments from XMCD spectra provide only the $d$-shell
contributions to the total spin and orbital moments. 
In Fig.~1 these values are compared to  the calculated total spin and orbital 
 moments.  However, as it follows from the calculations,  the
$s$- and $p$-contributions to the total spin moments
constitute at most 5\%. The corresponding contributions to the total
orbital moments are found even smaller and thus could be neglected.

Among other advantages, the SPR-KKR  method utilized in the present work
can straightforwardly combine the Coherent Potential Approximation (CPA)
theory describing disordered alloys with  the DMFT scheme. The latter is
illustrated for the bcc Fe$_{x}$Co$_{1-x}$ disordered alloys.
\begin{figure}
\includegraphics[width=0.45\textwidth,angle=0]{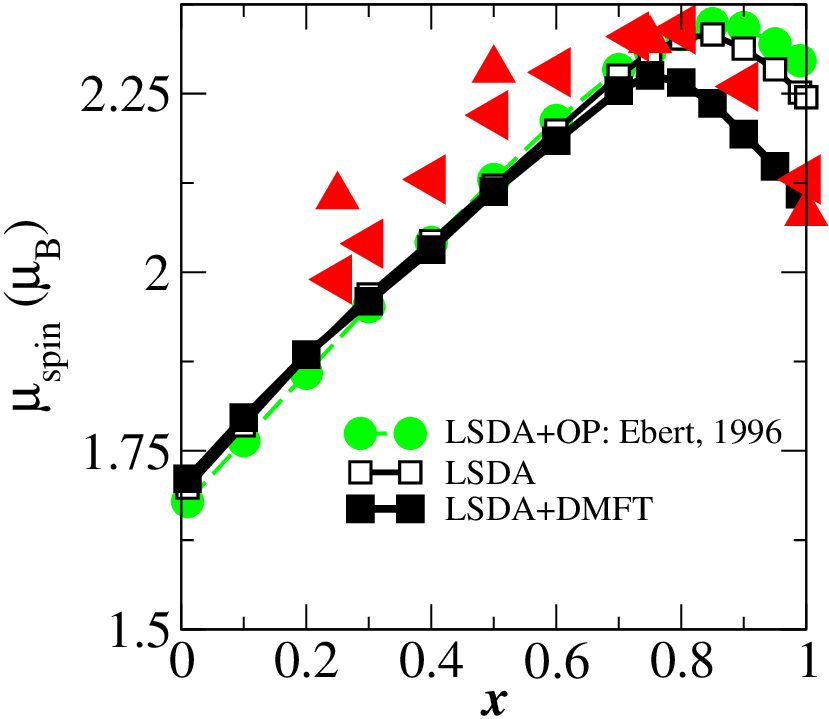}~
\includegraphics[width=0.45\textwidth,angle=0]{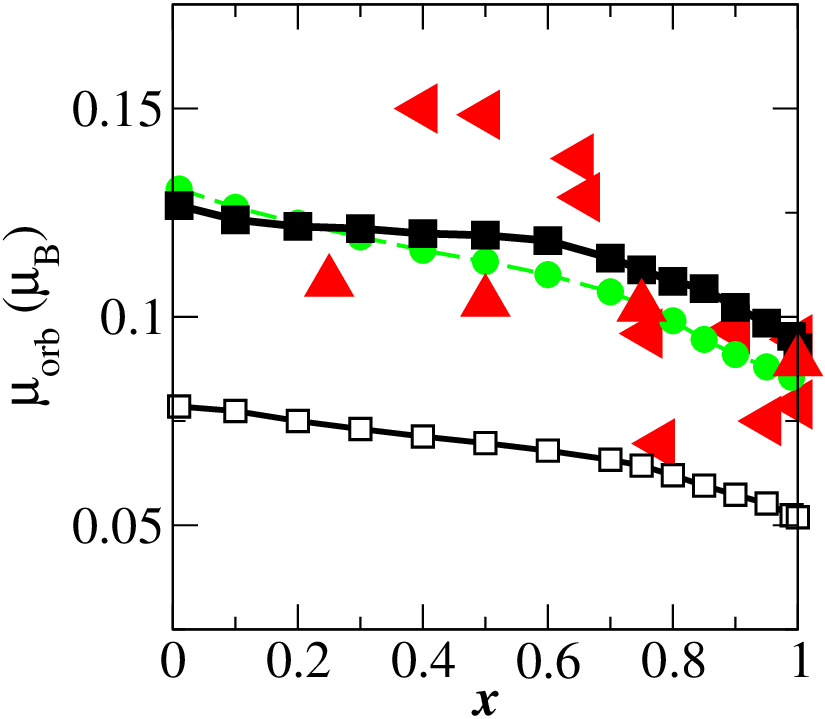}
\caption{(color online) Spin (left panel) and orbital (right panel)
  total magnetic moments 
of disordered bcc Fe$_x$Co$_{1-x}$ alloys  calculated as a function of
 Fe concentration $x$ via LSDA+DMFT (black line, filled squares), 
compared to present plain LSDA (black line, open squares), LSDA+OP
calculations \cite{EB96} (green line, filled circles),  
and experimental data (red triangles) taken from Refs.~\cite{Ste84}.
The corresponding DMFT parameters  are $U_{\rm Fe}=2$~eV, 
$U_{\rm Co}=3$~eV and  $J_{\rm Fe}=J_{\rm Co}=0.9$~eV. In nature 
the bcc structure  exists only for $x>0.25$.} 
\label{FeCo-moments}
\end{figure}
As one can see from Fig.~4, while the spin
magnetic moments for all approaches agree rather well,
LSDA+DMFT considerably improves the orbital moments in comparison
to plain LSDA calculations in a way similar to the result obtained
by Ebert and Battocletti using the LSDA+OP combined with the CPA
\cite{EB96}. Also in contrast to both the LSDA and LSDA+OP
calculations, a more pronounced agreement with experimental spin
magnetic moments is achieved  by LSDA+DMFT within the Fe-rich area of
concentrations.

As a further example we consider the unquenching of the orbital moment
of 3$d$-TM impurities embedded in Au. Tab.~1 illustrates the  results of LSDA+DMFT 
calculations compared to the experimental orbital to spin moment
ratios.
\begin{table}
\caption{Orbital to spin ratios of total magnetic moments 
of diluted Fe and Co impurities embedded in fcc Au host calculated  
via LSDA+DMFT compared with present plain LSDA  and experimental data
\cite{BSS+04}. The corresponding DMFT parameters are 
 are $U_{\rm Fe}=2$~eV, $U_{\rm Co}=3$~eV and 
$J_{\rm Fe}=J_{\rm Co}=0.9$~eV.}
\begin{center}
\begin{tabular}{l|cccc}
                       &  LSDA & LSDA+DMFT &LSDA+OP\cite{BSS+04}& Exp. \\ \hline
Fe$_{0.008}$Au$_{0.992}$  & 0.007 & 0.018  & 0.098 &0.034 \\
Co$_{0.015}$Au$_{0.985}$  & 0.109 & 0.345  & 0.7&0.336 
\end{tabular}
\end{center}
\end{table}
For the case of a Co impurity we have reached a drastic improvement 
when comparing to previous OP studies
\cite{BSS+04,Fro04,Sar06}; the latter gives a ratio of orbital
to spin magnetic moments of about $0.7$  whereas the ratio calculated 
in the present work is very close to the experimental 
value $0.35$. For the case of Fe the agreement
with experiment is not perfect but still reasonable. 

Here we want to mention
that no relaxation of the lattice near the impurity was considered. As reported
in Ref.~\cite{SOR+05} for the case of an Co impurity in Au
the additional lattice relaxation (about
2\%) leads to a about 5\% decrease for the spin and 28\% decrease for
the orbital magnetic moment of the impurity atom, leading to a 
25\% reduction for the orbital to spin moment ratio.
Assuming corresponding changes for our results gives
good agreement with experiment. Thus, although the
lattice relaxation around the impurity site might be an important
factor a complete description of the magnetic properties 
of impurities require first of all a satisfying treatment of  
 correlation effects.

Summarizing  the  results, we  emphasize  that  the presented  LSDA+DMFT
scheme which has proven already its efficiency in the description of the
spectral  properties   of  3$d$-TMs   has  also  greatly   improved  the
description of the orbital  magnetic moments for pure transition metals,
their alloys and impurities in noble metal host.

\section{Acknowledgements}
This work was supported by the Deutsche Forschungsgemeinschaft within
the priority program ``Moderne und universelle first-principles-Methoden
f\"ur Mehrelektronensysteme in Chemie und Physik'' (SPP 1145/2).


\bibliographystyle{apsrev}                 

\end{document}